\documentclass[aps,prb,amsmath,amssymb,twocolumn,superscriptaddress]{revtex4-2}
\usepackage{graphicx}
\usepackage{subfigure}
\usepackage{adjustbox}
\usepackage{bm}
\usepackage{color}
\usepackage{braket}
\usepackage{standalone}
\usepackage{multirow}
\usepackage{tikz}
\usepackage{mathrsfs}
\usepackage{dsfont}
\usepackage{times}

\usepackage[colorlinks,bookmarks=true,citecolor=blue,linkcolor=blue,urlcolor=blue]{hyperref}
\newcommand*{\id}{{\normalfont\hbox{1\kern-0.15em \vrule width .8pt depth-.5pt}}}

\begin{document}

\title{Finite-rate quench in disordered Chern and $Z_2$ topological insulators}
\author{Sheng-Nan Du}
\affiliation{Zhejiang Institute of Modern Physics, Zhejiang University, Hangzhou 310058, China}
\author{Zhao Liu}
\email[Contact author: ]{zhaol@zju.edu.cn}
\affiliation{Zhejiang Institute of Modern Physics, Zhejiang University, Hangzhou 310058, China}
\affiliation{Zhejiang Key Laboratory of Micro-Nano Quantum Chips and Quantum Control, School of Physics, Zhejiang University, Hangzhou 310027, China}
%\date{\today}

\begin{abstract}
We study the quantum quench of a finite rate across topological quantum transitions in two-dimensional Chern and $Z_2$ topological insulators. We choose the representative Haldane model and the Kane-Mele model to investigate the behavior of excitation density generated by the quench and the impact of disorder on it. For the Haldane model, as long as the spectral gap is not closed by disorder, we find the excitation density at the end of the quench obeys the power-law decay with decreasing quench rate, and the power is consistent with the prediction of the Kibble-Zurek mechanism. By contrast, the Kibble-Zurek scaling of excitation density is absent in the Kane-Mele model once disorder is switched on, which we attribute to the emergence of a disorder-induced gapless region. In particular, the anti-Kibble-Zurek behavior of excitation density, namely, larger excitation density at slower quench, is observed at suitable model parameters. Moreover, we demonstrate that particle's onsite occupation can be used as a local measurable quantity to probe the breakdown of adiabatic evolution. The difference of onsite occupation between the time-evolved state and instantaneous ground state at the end of the quench can successfully capture the key features of excitation density for both the Haldane and Kane-Mele models under periodic and more realistic open boundary conditions, thus facilitating the experimental characterization of the quench dynamics in these models. 
\end{abstract}
\maketitle

\section{Introduction}\label{introduction}
The quantum adiabatic theorem~\cite{Born,Kato} is a useful tool for understanding the evolution of quantum systems subject to external driving.  For a system initially prepared in the ground state of a time-dependent Hamiltonian, it will track the instantaneous ground state, provided that the Hamiltonian varies sufficiently slowly and an energy gap separating the ground state from instantaneous excited levels exists at any time during the evolution. The success of quantum adiabatic evolution is crucial for a number of schemes in quantum simulation and computation, including, for example, adiabatic quantum computation~\cite{RevModPhys.90.015002}, quantum annealing~\cite{Annealing}, and anyon braiding~\cite{RevModPhys.80.1083}.

The adiabatic evolution is unable to account for the behavior of the system when the relaxation time, which is inversely proportional to the instantaneous energy gap of the Hamiltonian, becomes larger than the timescale of the Hamiltonian's variation. In the thermodynamic limit, no matter how slowly the Hamiltonian varies, the breakdown of adiabatic evolution is unavoidable if the system is driven across the critical point of a quantum phase transition, where the energy gap protecting the instantaneous ground state vanishes. In finite systems with a fixed variation speed of the Hamiltonian, the probability of jumping to excited levels is the highest at the point with the minimal instantaneous energy gap. The failure of the system to adiabatically follow external driving provides a mechanism of dynamical formation of excitations. 

In the context of slow linear quench across the critical point of a second-order quantum phase transition, the Kibble-Zurek (KZ) mechanism provides a theoretical framework for understanding the emergence of excitations. While the KZ mechanism was initially proposed for cosmological evolution and thermodynamic second-order phase transitions involving symmetry breaking~\cite{0-Kibble_1976,0-Zurek_1985}, it has been generalized to the setting of quantum phase transitions~\cite{PhysRevLett.95.245701,54-Zurek_2005,PhysRevB.72.161201,PhysRevB.81.012303,PhysRevB.86.064304,KZreview,DziarmagaReview,RevModPhys.83.863,Fischer_PhysRevLett2006,Fischer_PhysRevLett2007,Fischer_PhysRevA2008,Fischer_PhysRevD2010,Uhlmann_2010}. For a $d$-dimensional system, a power-law relation $n_{\rm ex}\sim \tau^{-d\nu/(1+z\nu)}$ is predicted between the density of dynamical excitations $n_{\rm ex}$ and the quench rate $\tau$, where $\nu$ is the correlation length critical exponent and $z$ is the dynamic critical exponent of the phase transition. The KZ behavior has been experimentally tested in a variety of systems, ranging from superfluids~\cite{21-Mathey_2010, 22-Chesler_2015, 23-Ko_2019} and superconductors~\cite{PhysRevLett.89.080603,24-Monaco_2006, 25-Garaud_2014, 26-Sonner_Zurek_2015, 27-Defenu_2019} to quantum Ising models~\cite{28-Dziarmaga_2006, 29-Silvi_2016, 30-Kolodrubetz_2012, 31-Sinha_2020}, multiferroics~\cite{PhysRevLett.108.167603,PhysRevX.2.041022,PhysRevX.7.041014}, Coulomb crystals~\cite{32-Mielenz_2013, 33-Ulm_2013, 34-Pyka_2013}, Bose gases~\cite{35-Navon_Science_2015, 36-Chomaz_2015, 37-Beugnon_2017, 38-Goo_2021, 39-Navon_natphys_2021}, Bose–Einstein condensates~\cite{BEC}, Rydberg simulators~\cite{Rydberg}, and cosmological scenarios~\cite{40-Bunkov_1998, 41-Eltsov_2000, 42-Casado_2007}.

Topological phases of matter have sparked great interest since being discovered. Numerous efforts have been made with the aim of understanding the dynamical response of topological systems to external driving~\cite{PhysRevLett.115.236403,PhysRevB.94.155104,PhysRevLett.117.126803,PhysRevLett.117.235302,PhysRevLett.121.090401,PhysRevLett.121.250601}. Remarkably, despite the absence of symmetry breaking in transitions involving topological phases, the KZ scaling of excitations and real-space topological index have been reported in the bulk of several systems evolving across topological phase transitions~\cite{PhysRevB.78.045101,43-Bermudez_2009, 45-Dutta_2010,44-Lee_Choi_2015, 52-KunYang_PRB_2018,46-Ulcakar_PRB_2018, 47-Ulcakar_PRB_2019, 48-Ulcakar_PRL_2020,LiFuxiang2022, LiFuxiang2024}. However, the characteristic edge states of topological phases destroy the KZ scaling, resulting in the anti-KZ behavior of dynamical excitations~\cite{PhysRevX.2.041022,antiKZ_PhysRevLett2016,antiKZ_PhysRevB2017,anti_PhysRevA2021,antiKZ_PhysRevB2021,antiKZ_PhysRevLett2020} in systems with open boundary conditions~\cite{52-KunYang_PRB_2018}.

In this work, we conduct a further investigation of the quench dynamics across topological phase transitions in two spatial dimensions (2D). There are two issues in which we are particularly interested: (i) how robust is the previously reported KZ scaling in topological phase transitions against disorder; (ii) what measurable quantity can be used to detect the breakdown of adiabatic evolution and capture the dynamical features (such as the KZ and anti-KZ behavior) of excitation density. To be concrete, we consider a linear quench in two paradigmatic 2D lattice models, the Haldane model~\cite{1-Haldane_1988} and the Kane-Mele model~\cite{2-KaneMele_QSH_2005, 3-KaneMele_Z2_QSH_2005}, which can support Chern insulators and $Z_2$ topological insulators, respectively. We incorporate disorder in both models. We drive the system from a trivial phase to the topological phase by tuning specific model parameters linearly with time. We use the overlaps between time-evolved states and instantaneous eigenstates of the single-particle Hamiltonian to define the excitation density generated by the quench. The critical exponents of the pertinent static phase transitions are carefully extracted in the presence of disorder. Under periodic boundary conditions (PBCs), we find robust KZ scaling of excitation density in the disordered Haldane model, which persists until the spectral gap of the system is closed by very strong disorder. This result hence generalizes the KZ behavior reported for topological systems with tiny disorder~\cite{LiFuxiang2022,48-Ulcakar_PRL_2020}. However, the situation is more complicated in the Kane-Mele model. In that case, we identify a disorder-induced gapless region between the topological and trivial phases. When the system is driven through this region, the KZ scaling of excitation density is violated. In particular, the excitation density may grow with decreasing quench rate at suitable model parameters, showing the anti-KZ behavior. To seek an experimental scheme to detect these dynamic features, we consider the particle's occupation on a specific lattice site, using the difference between the occupation obtained from the time-evolved state and the instantaneous ground state to characterize the breakdown of the adiabatic evolution. For both the Haldane and Kane-Mele models under PBCs, we find this onsite occupation difference captures the key features of excitation density's dependence on quench rate, including the presence and absence of the KZ scaling. More interestingly, even under open boundary conditions (OBCs), evaluating this onsite occupation difference in the bulk can well reproduce the results under PBCs. Therefore, it can be used as a local quantity to characterize the finite-rate quench dynamics of realistic systems. We expect similar results for other models satisfying certain conditions. 

Our paper is organized as follows. In Sec.~\ref{theory and model}, we review the KZ theory and introduce the Haldane and Kane-Mele models studied in this work. We also present the methods used to extract the critical exponents of topological phase transitions and simulate the time evolution. The definition of excitation density will be given. In Sec.~\ref{disorder effect}, we investigate linear quench across topological phase transitions in disordered Haldane and Kane-Mele models under PBCs. We compare the prediction of KZ theory with the numerical simulation of the linear quench. In Sec.~\ref{electron occupation}, we present the dynamics of particle's onsite occupation under both PBCs and OBCs. We conclude with some outlooks in Sec.~\ref{c_and_o}. More numerical data will be provided in the Appendix.

\section{Model and method}\label{theory and model}
In this section, we review the basic idea of the KZ theory and introduce the Haldane model and the Kane-Mele model studied in this paper. Then we present our methods which are used to simulate the time evolution and examine the KZ prediction. 

\subsection{Kibble-Zurek theory}\label{KZ_theory}
We consider a quantum system whose Hamiltonian depends on a parameter $\lambda$ and exists a second-order quantum phase transition at the critical point $\lambda=\lambda_c$. At some time $t<0$, the system is initially prepared as the ground state of the Hamiltonian with $\lambda$ far from the critical point. Then it is driven out of equilibrium by varying $\lambda$ with time. Under the assumption of linear quench, we can define $ \lambda(t)/\lambda_c-1=t/\tau$, such that $|t|$ measures the temporal distance to the critical point which is reached at $t=0$. The quench rate is given by $1/\tau$. We terminate the quench at some time $t>0$, corresponding to $\lambda$ far from the critical point on the other side.

In the thermodynamic limit, the adiabatic evolution of the system is impossible at $t=0$ no matter how slow the quench is. This is because the energy gap $ \Delta_{\rm s} $ above the instantaneous ground state vanishes as
\begin{equation}\label{exci1}
\Delta_{\rm s}(t)\sim\bigg|\dfrac{\lambda(t)}{\lambda_c}-1\bigg|^{z\nu}=\left|\frac{t}{\tau}\right|^{z\nu}
\end{equation}
when the critical point is approached. For a fixed quench rate, the switch from the adiabatic to the diabatic regimes occurs if the relaxation time $\tau_R$, which is determined by the energy gap via $\tau_R=\hbar/\Delta_{\rm s}$, grows to be comparable to the temporal distance $|t|$ from the critical point. The instant $\hat{t}$ at which this happens scales with $\tau$ as 
\begin{equation}\label{exci2}
\hat{t}\sim\tau^{z\nu/(1+z\nu)}.
\end{equation}
In this setup, the tracking of the instantaneous ground state ceases at time $t=-\hat{t}$, and evolution from the ``frozen out'' state restarts at $t= +\hat{t}$. The separation of the evolution into adiabatic and diabatic regimes is the essence of the KZ theory.

The breakdown of the adiabatic evolution creates excitations in the system. In the standard KZ mechanism for a thermodynamic second-order phase transition from a high-symmetry phase to a broken-symmetry phase, there are $O(1)$ excitations (defects) per $\xi$, where $\xi$ is the correlation length at time $t=-\hat t$. The density of excitations is then given by $ n_{\rm ex} \sim\xi^{-d}$ for a $d$-dimensional system. Adopting this argument in second-order quantum phase transitions involving symmetry breaking~\cite{54-Zurek_2005}, and using 
\begin{equation}\label{exci3}
\xi\sim\bigg|\dfrac{\lambda(-\hat t)}{\lambda_c}-1\bigg|^{-\nu}=\left|\frac{\hat t}{\tau}\right|^{-\nu},
\end{equation}
one can expect that the excitation density scales as
\begin{equation}\label{exci_tau}
n_{\rm ex}\sim\xi^{-d}\sim\left|\frac{\hat t}{\tau}\right|^{d\nu}\sim\tau^{-\frac{d\nu}{1+z\nu}}.
\end{equation}
Remarkably, the scaling of excitations generated during the dynamic process is determined by the critical exponents of the static quantum phase transition. 

For generic second-order quantum phase transitions, especially those without symmetry breaking, it is more reasonable to use the standard Landau-Zener (LZ) formula~\cite{LZ} to evaluate the excitation density generated by the quench if the model can be expressed as a collection of two-level systems~\cite{PhysRevLett.95.245701,54-Zurek_2005,PhysRevB.72.161201,PhysRevLett.100.077204,PhysRevB.78.045101,PhysRevB.78.144301,PhysRevB.106.224302,52-KunYang_PRB_2018}. In many cases, the LZ formula gives the same scaling of the excitation density with the quench rate as the KZ prediction, despite a different prefactor~\cite{PhysRevLett.95.245701,54-Zurek_2005,PhysRevB.78.045101}. However, it has also been reported in several systems that the KZ exponent in Eq.~(\ref{exci_tau}) should be generalized~\cite{PhysRevLett.100.077204,PhysRevB.78.045101,PhysRevB.78.144301,PhysRevB.106.224302,52-KunYang_PRB_2018,LiFuxiang2024}. In particular, when the quench takes a $d$-dimensional system with translation invariance through a $d-m$ dimensional gapless surface in the momentum space, the excitation density is found to scale as $\tau^{-m\nu/(1+z\nu)}$~\cite{PhysRevB.78.045101,PhysRevLett.100.077204,PhysRevB.106.224302}. Quench across a gapless point corresponds to $m=d$, which returns to the standard KZ prediction.

\subsection{Haldane model}\label{haldane_model}
The Haldane model is a paradigm model for the integer quantum Hall effect without Landau levels, dubbed Chern insulator~\cite{1-Haldane_1988}. It is defined for spinless fermions on a honeycomb lattice, consisting of real hopping of fermions between nearest-neighbor (NN) sites and complex hopping between the next-nearest-neighbor (NNN) sites. The nontrivial phase in the complex NNN hopping breaks the time-reversal symmetry, giving rise to the possibility of nonzero band Chern number. There are also opposite energy offsets on the two sublattices of the honeycomb, which further breaks the inversion symmetry. In this work, we incorporate onsite disorder, and add extra real next-next-nearest-neighbor (NNNN) hopping to flatten the band and reduce the finite-size effect in the presence of disorder~\cite{Priest2014}. 

With all ingredients above, the single-particle Hamiltonian of the Haldane model is given as
\begin{eqnarray}\label{Haldane}
H =&-&t_{1}\sum_{\langle i, j\rangle} c_{i}^{\dagger}c_{j} + t_{2}\sum_{\langle\langle i, j\rangle\rangle} e^{iv_{ij}\phi}c_{i}^{\dagger}c_{j} \nonumber \\
&+& t_{3}\sum_{\langle\langle\langle i, j\rangle\rangle\rangle} c_{i}^{\dagger}c_{j} 
   + \sum_{i}(M_i+w_i)c_{i}^{\dagger}c_{i} ,
\end{eqnarray}
where $ c_{i}^{\dagger} $ is the fermionic creation operator on the lattice site $i$, and $ \langle\cdot\rangle$, $\langle\langle\cdot\rangle\rangle$, and $\langle\langle\langle\cdot\rangle\rangle\rangle $ denote the NN, the NNN, and the NNNN pairs of sites, respectively. The phase of the NNN complex hopping depends on $v_{ij}\equiv(2/\sqrt 3) (\hat{\bm d}_{1}\times \hat{\bm d}_{2})\cdot \hat{z}=\pm 1$, where $ \hat{\bm d}_{1} $ and $ \hat{\bm d}_{2} $ are the unit vectors along the two bonds that connect site $ i $ to $ j $. $M_i=\pm M$ is the staggered potential alternating sign between the $A$ and $B$ sublattices of the honeycomb. $ w_{i} $ is the onsite disorder potential uniformly distributed in $ [-W, W] $. We set $t_1=1$ throughout the paper as the energy unit for the Haldane model.

In the absence of disorder, the Haldane model has two energy bands. Depending on the values of parameters, these two bands can carry either zero or nonzero Chern numbers, corresponding to trivial and Chern insulator phases, respectively~\cite{1-Haldane_1988}. The phase transition between them, accompanied by the closing of the band gap, can be achieved by tuning specific parameters like $M$, $\phi$, and $t_2$. When disorder is present, the Chern insulator can be distinguished from the trivial phase by the quantized Hall conductance at half filling.

\subsection{Kane-Mele model}\label{KM_model}
Kane and Mele generalized the Haldane model to time-reversal symmetric spin-$1/2$ fermions on a honeycomb lattice~\cite{3-KaneMele_Z2_QSH_2005}. The Hamiltonian of the model is
\begin{eqnarray}\label{Kane-Mele}
H = &-&t\sum_{{\langle i,j\rangle},\sigma}c^{\dagger}_{i\sigma}c_{j\sigma}+\sum_{i,\sigma}\left( M_{i}+w_{i}\right) c^{\dagger}_{i\sigma}c_{i\sigma} \nonumber \\
 & +&i\lambda_{\rm S}\sum_{\langle\langle i,j\rangle\rangle}\sum_{ \sigma_{1},\sigma_{2}}v_{ij}c^{\dagger}_{i\sigma_{1}}s^{z}_{\sigma_{1}\sigma_{2}}c_{j\sigma_{2}}\nonumber  \\
 & +&i\lambda_{\rm R}\sum_{\langle i,j\rangle}\sum_{\sigma_{1},\sigma_{2}}c^{\dagger}_{i\sigma_{1}}\left(\bm s_{\sigma_{1}\sigma_{2}}\times \hat{\bm d}_{ij}\right)_{z}c_{j\sigma_{2}},                                                                                                                                                                                                         
\end{eqnarray}
where $ c_{i\sigma}^{\dagger} $ creates a fermion with a $z$-component spin $\sigma=\uparrow$ or $\downarrow$ on lattice site $i$. The first three terms correspond to two independent Haldane models of opposite $z$ spins with $\phi=\pm\pi/2$, which are the time-reversal conjugate of each other. Note that the disorder potential $w_i$ is identical for two spin species, thus preserving time-reversal symmetry. The operator vector ${\bm s}\equiv (s^x,s^y,s^z)$ contains the three Pauli matrices. The $\lambda_{\rm S}$ term can originate from the intrinsic spin-orbit coupling (SOC) of graphene. The last term represents the realistic Rashba SOC, which breaks the conservation of the $z$-component of spin. We set $t=1$ throughout the paper as the energy unit for the Kane-Mele model.

In the absence of disorder, the Kane-Mele model has four energy bands. When the lowest two bands are occupied, the system is an $Z_2$ topological insulator under suitable parameters~\cite{3-KaneMele_Z2_QSH_2005}. The $Z_2$ topological insulator belongs to a different topological class from the Chern insulator. It is characterized by the $Z_2$ topological index or spin Chern number~\cite{PhysRevLett.97.036808,Prodan_2011} instead of the usual Chern number. Phase transition from the $Z_2$ topological insulator to a trivial insulating phase can be driven by varying model parameters like $M$, $\lambda_{\rm S}$, and $\lambda_{\rm R}$. When disorder is present, the topological phase can be distinguished from the trivial phase by the nonzero integer spin Chern number at half filling.

\subsection{Method}
\label{sec::method}
Throughout this work, we focus on half filling for both the Haldane and Kane-Mele models. When the system is in the static ground state, the low-energy half of all single-particle levels are occupied. For clean systems, the lower one band (two bands) in the momentum space is fully occupied for the Haldane (Kane-Mele) model. Due to the noninteracting nature of the system, the many-body wavefunction is simply a Slater determinant of all occupied single-particle states.  

We use the collapse of the spectral gap above the ground state to identify the critical points of the static topological phase transitions induced by tuning model parameters. For our noninteracting systems, the spectral gap is just the energy difference between the highest occupied and the lowest empty single-particle levels. We choose $t_2$ and $\lambda_{\rm S}$ as the tuning parameters for the Haldane and Kane-Mele models, respectively. The scaling behavior of the spectral gap in finite systems near the critical point can provide an estimation of the critical exponents $\nu$ and $z$ characterizing the phase transition. For a system with length scale $L$, the spectral gap exactly at the critical point $\lambda_c$ of a second-order quantum phase transition scales with the system size as
\begin{equation}\label{Sgap_L_z}
\Delta_{\rm s}(\lambda_c, L)\sim L^{-z},
\end{equation}
which exponentially vanishes in the thermodynamic limit $L\rightarrow \infty$. On the other hand, the spectral gap in the thermodynamic limit should show a power-law relation with the deviation from the critical point:
\begin{equation}\label{Sgap_L_z_nu}
\Delta_{\rm s}(\lambda, L\rightarrow\infty)\sim|\lambda-\lambda_c|^{z\nu}.
\end{equation}
Combing Eqs.~(\ref{Sgap_L_z}) and (\ref{Sgap_L_z_nu}), we expect the scaling behavior
\begin{equation}\label{rescaledSgap_L_z_nu}
\Delta_{\rm s}(\lambda,L)\sim L^{-z} f[L^{1/\nu}(\lambda-\lambda_{c})]
\end{equation}
for the spectral gap in a finite system near the critical point, where $f$ is a universal function. Data collapsing of the spectral gaps of different system sizes with suitable choices of $z$, $\nu$, and $\lambda_c$ can return the estimated values of $z$ and $\nu$, based on the which we can get the exponent in the KZ prediction (\ref{exci_tau}) for the scaling of excitation density. This method should apply for both clean and disordered systems.

To compare the actual behavior of the dynamical excitation density with the KZ prediction, we numerically simulate the time evolution of the system driven by the time-dependent linear variation of model parameters. For numerical convenience, we shift the starting point of the time to $t=0$. For a specific model parameter $\lambda$, we fix its initial value $\lambda_i$ and final value $\lambda_f$, and vary it with time as $\lambda(t)=\lambda_i+(\lambda_f-\lambda_i)t/\tau$, such that $\lambda=\lambda_i$ at $t=0$ and $\lambda=\lambda_f$ at $t=\tau$. We assure finite static spectral gaps at both $\lambda_i$ and $\lambda_f$. As the system is noninteracting, its time evolution is determined by the evolution of each single-particle eigenstate of the Hamiltonian. We discretize the whole duration of the quench from $t=0$ to $t=\tau$ into $N$ steps. In each time interval $\Delta t=\tau/N$ between $t=(n-1)\Delta t$ and $t=n\Delta t$, the time evolution corresponding to the $m$th single-particle eigenstate can be approximated as  $ |\varphi_m(n\Delta t)\rangle={\rm e}^{-i\Delta t H[\lambda((n-1)\Delta t)]}|\varphi_m((n-1)\Delta t)\rangle $. Initially we have $|\varphi_m(t=0)\rangle=|\psi_m(\lambda_i)\rangle$, where $|\psi_m(\lambda_i)\rangle$ is the $m$th instantaneous single-particle eigenstate of the Hamiltonian at $\lambda=\lambda_i$. We then define the excitation density at time $t$ as
\begin{equation}\label{excitation_numerical}
n_{\rm ex}(t)=1-\frac{1}{N_s/2}\sum_{m,n=0}^{\frac{N_s}{2}-1}|\langle\psi_m(\lambda(t))|\varphi_n(t)\rangle |^{2},
\end{equation}
where $N_s$ is the total number of single-particle levels (sorted in ascending order) for the finite system. Plotting $n_{\rm ex}$ at the end of the quench as a function of the quench rate $\tau$ can test the validity of the KZ prediction. 

In the presence of disorder, the spectral gap, the time-evolution, and the excitation density should be calculated for multiple random realizations of the disorder potential $\{w_i\}$. For each sample, the configuration of the disorder potential is fixed during the time evolution. The final results presented in this work have been averaged over sufficient disorder samples to get reasonably small error bars. Simulating slower quench is more difficult, because larger $\tau$ means longer evolution and requires denser discretization of the time. Limited by the difficulty of simulating slow quench and the necessity of considering various disorder configurations, the largest $\tau$ we reach in this work is typically several hundreds.  

\section{Disorder effect on the scaling of excitation density}\label{disorder effect}
In this section, we study the effect of disorder on the excitation density after the quench. In particular, we examine the robustness of previously reported KZ scaling of excitation density in topological phase transitions of almost clean systems with respect to increasing disorder strength. We consider finite systems with $L$ unit cells in each of the two primitive directions of the honeycomb lattice. We impose PBCs in this section, i.e., the system is put on the torus. While it is possible to utilize the momentum conservation to simplify the simulation of clean systems, we have to work in the real space if disorder is present, which breaks the translation symmetry of the system. 

\begin{figure}
	    \centering
         \includegraphics[width=0.9\linewidth]{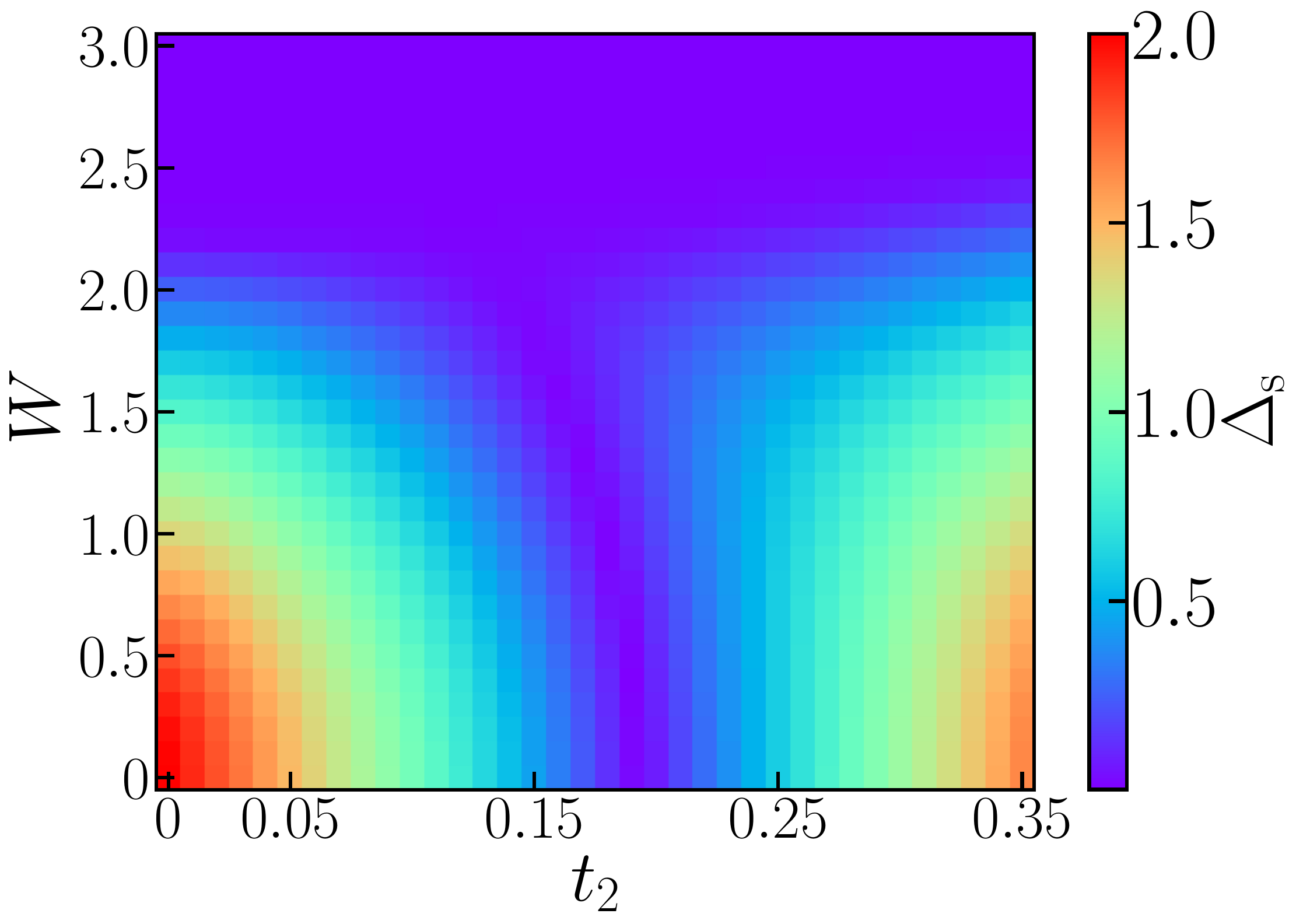}
         \caption{The spectral gap $ \Delta_{\rm s} $ for the Haldane model with $ L=60 $ as a function of $ t_{\rm 2} $ and $ W $. The data are averaged over 100 disorder configurations at each parameter point. Other model parameters are chosen as $ M=1$, $ \phi=\pi/2$ and $ t_{3}=1/3 $. For $ W<2 $, the spectral gap develops a minimum at a critical point $ t_{2c} $ when $ t_{2} $ is tuned from $ 0 $ to $ 0.35 $. }
          \label{1-Sgap_Haldane}
 \end{figure} 
 
 \begin{figure*}
	    \centering
         \includegraphics[width=\linewidth]{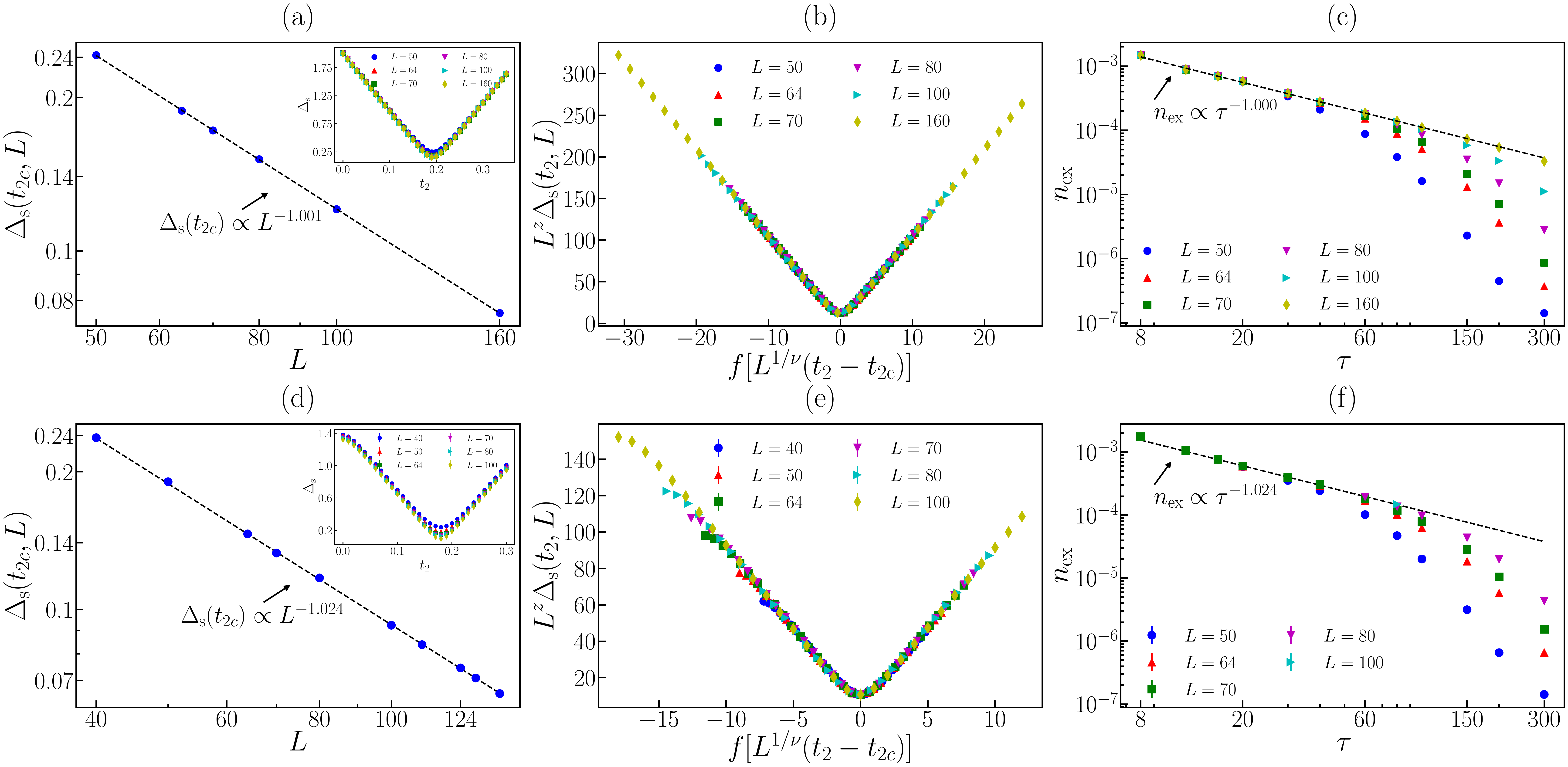}
         \caption{(a) The spectral gap at the critical point $ t_{2c}=0.19$ for the clean Haldane model of different system sizes $ L $ (on a log-log scale), giving $ z\approx1.001 $. The inset shows the evolution of the spectral gap with $ t_{\rm 2} $, with the minimum at $ t_2\approx 0.19$. (b) The rescaled variable $L^z \Delta_{\rm s}(t_2,L)$ versus $L^{1/\nu}(t_2-t_{2c})$ for the clean Haldane model. Data of different $ L $ collapse to the same curve under the choice of $ \nu=1 $. (c) The dependence of the final excitation density $ n_{\rm ex} $ after the linear quench on the quench rate $ 1/\tau $ (on a log-log scale) for the clean Haldane model. For large enough systems, $ n_{\rm ex} $ shows a power-law dependence $ n_{\rm ex}\sim\tau^{-\alpha}$ with $ \alpha\approx 1.000 $. (d)-(f) The same quantities as in (a)-(c), but for the disordered Haldane model with $ W=1 $. We get $ t_{{\rm 2}c}\approx 0.18$, $ z\approx1.024 $, $ \nu\approx1 $, and $ \alpha\approx 1.024 $. In both the clean and disordered cases, we fix $ M=1$, $ \phi=\pi/2$ and $ t_{3}=1/3 $, and drive the quench by varying $ t_{\rm 2} $ from $ 0.1 $ to $ 0.3 $.}
         \label{2-Sgap_rescaleSgap_Exci_clean_W_1_Haldane}
 \end{figure*}

\subsection{Haldane model}\label{Haldane model}
We study the Haldane model with fixed parameters $ M=1$, $ \phi=\pi/2$, and $ t_{3}=1/3 $. In the absence of disorder (disorder strength $W=0$), the spectral gap at half filling in the thermodynamic limit is closed at the critical $t_2$, where the topological phase transition occurs and the Chern number of the lower band changes between 0 and 1. For finite systems, the vanishing of the spectral gap is replaced by a minimum. The spectral gap of $L=60$ at half filling as a function of $ t_{2} $ and $ W $ is shown in Fig.~\ref{1-Sgap_Haldane}. The minimum of the spectral gap still exists at a specific value of $t_2$ in disordered finite systems until $W\approx 2$. By calculating the Hall conductance, we confirm that the transition from a trivial insulating phase to the Chern insulator phase with the increasing of $t_2$ persists for $ W<2 $.

Before we investigate the quench dynamics in disordered systems, we first use the clean limit to examine the validity of our numerical method outlined in Sec.~\ref{sec::method}. In this case, the dependence of the spectral gap $\Delta_{\rm s}(t_{2c},L)$ at the critical point $t_2=t_{2c}$ on the system size $L$ is shown in Fig.~\ref{2-Sgap_rescaleSgap_Exci_clean_W_1_Haldane}(a). We choose $ t_{2c}=0.19 $, based on the location of the minimal gap for different system sizes, as displayed in the inset of Fig.~\ref{2-Sgap_rescaleSgap_Exci_clean_W_1_Haldane}(a). As $\Delta_{\rm s}(t_{2c},L)$ linearly decays with $L$ [Fig.~\ref{2-Sgap_rescaleSgap_Exci_clean_W_1_Haldane}(a)], we conclude $z=1$ for the phase transition. We further analyze the finite-size spectral gap $\Delta_{\rm s}(t_{2},L)$ near $t_2=t_{2c}$. By plotting the rescaled variable $L^z \Delta_{\rm s}(t_2,L)$ versus $L^{1/\nu}(t_2-t_{2c})$, we find all data of various system sizes collapse onto a single curve if we set $\nu=1$  [Fig.~\ref{2-Sgap_rescaleSgap_Exci_clean_W_1_Haldane}(b)]. Our numerical extraction of $z$ and $\nu$ matches the theoretically known $z=\nu=1$ for Dirac fermions. With these values of $z$ and $\nu$, the KZ theory predicts $n_{\rm ex}\sim \tau^{-1}$. This prediction is consistent with our numerical simulation of the quench. We display the typical quench result in Fig.~\ref{2-Sgap_rescaleSgap_Exci_clean_W_1_Haldane}(c), in which $ t_{2} $ linearly varies with time from $ t_2=0.1 $ to $ t_2=0.3 $. The final value of the excitation density defined in Eq.~(\ref{excitation_numerical}) clearly scales with $\tau$ as $n_{\rm ex}\sim \tau^{-1}$ for sufficiently large systems. Our results are consistent with those reported in Ref.~\cite{52-KunYang_PRB_2018}, in which the quench is driven by varying $M$. 

We also observe that the data points of $n_{\rm ex}$ for smaller systems start to deviate from the $\tau^{-1}$ scaling at smaller $\tau$ [Fig.~\ref{2-Sgap_rescaleSgap_Exci_clean_W_1_Haldane}(c)], due to which we have to discard some data points at large $\tau$ in the power-law fitting of $n_{\rm ex}$. This deviation is actually a finite-size effect, originating from the nonzero spectral gap near the critical point for any finite system. Therefore, the adiabatic evolution can be restored once the quench is sufficiently slow (large $\tau$). With a bigger spectral gap near the critical point, smaller systems return to adiabatic evolution at smaller $\tau$. We find that the excitation density scales as $\tau^{-2}$ at sufficiently large $\tau$ for finite systems (see Appendix~\ref{sec::largetau}), which is consistent with the prediction of adiabatic perturbation theory~\cite{PhysRevB.81.012303}. We see similar finite-size effects also in disordered systems. 

\begin{figure*}
	    \centering
         \includegraphics[width=\linewidth]{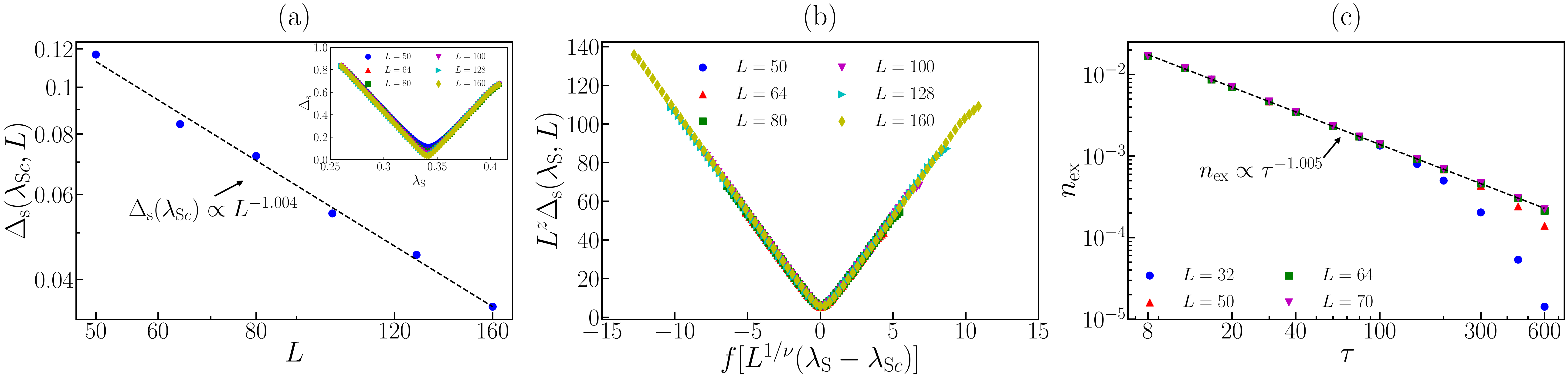}
         \caption{The spectral gap at the critical point $ \lambda_{{\rm S}c}=0.34$ for the clean Kane-Mele model of different system sizes $ L $ (on a log-log scale), giving $ z\approx1.004 $. The inset shows the evolution of the spectral gap with $\lambda_{{\rm S}} $, with the minimum at $ \lambda_{{\rm S}}\approx 0.34$. (b) The rescaled variable $L^z \Delta_{\rm s}(\lambda_{{\rm S}},L)$ versus $L^{1/\nu}(\lambda_{{\rm S}}-\lambda_{{\rm S}c})$ for the clean Kane-Mele model. Data of different $ L $ collapse to the same curve under the choice of $ \nu=1 $. (c) The dependence of the final excitation density $ n_{\rm ex} $ after the linear quench on the quench rate $ 1/\tau $ (on a log-log scale) for the clean Kane-Mele model of various system sizes. For large enough systems, $ n_{\rm ex} $ shows a power-law dependence $ n_{\rm ex}\sim\tau^{-\alpha}$ with $ \alpha\approx 1.005 $. We fix $M=0.5$ and $\lambda_{\rm R}=1 $ in our calculations, and drive the quench by varying $ \lambda_{\rm S} $ from $ 0.1 $ to $ 0.7 $.}
         \label{3-Sgap_rescaleSgap_Exci_clean_KaneMele}
 \end{figure*}

Having examined our numerical methods in the clean limit, we now turn to disordered systems. Due to the broken translation symmetry and the necessity to simulate various realizations of the disorder potential, we can only deal with smaller systems than in the clean limit. The results for a typical intermediate disorder strength $ W=1 $ are shown in Figs.~\ref{2-Sgap_rescaleSgap_Exci_clean_W_1_Haldane}(d)-\ref{2-Sgap_rescaleSgap_Exci_clean_W_1_Haldane}(f). In this case, the spectral gap $\Delta_{\rm s}(t_{2},L)$ suggests $t_{2c}\approx 0.18 $ and $ z\approx 1.024 $ [Fig.~\ref{2-Sgap_rescaleSgap_Exci_clean_W_1_Haldane}(d)]. We further extract $ \nu\approx1 $ from the data collapsing of $L^z \Delta_{\rm s}(t_2,L)$ versus $L^{1/\nu}(t_2-t_{2c})$, as shown in Fig.~\ref{2-Sgap_rescaleSgap_Exci_clean_W_1_Haldane}(e), which gives the KZ prediction $n_{\rm ex}\sim\tau^{-0.988}$. On the other hand, the numerical simulation of the quench dynamics suggests $n_{\rm ex}\sim\tau^{-1.024}$ [Fig.~\ref{2-Sgap_rescaleSgap_Exci_clean_W_1_Haldane}(f)], which is in good agreement with the KZ prediction. The estimated $z$, $\nu$, and the scaling exponent of the excitation density in the disordered case are also very close to those in the clean limit. We also obtain similar results for other disorder strength at $W<2$, as well as for the inverse quench driven by varying $ t_{\rm 2} $ from $ 0.3 $ to $ 0.1 $. This indicates that the KZ prediction is valid for the disordered Haldane model and the exponent of $\tau$ is robust before the spectral gap vanishes at strong disorder. We expect a further reduction of discrepancy from the clean limit if larger systems are reached in the simulation of dynamics, which is, however, computationally expensive for disordered systems and large $\tau$. 

\subsection{Kane-Mele model}\label{KM model}

We have demonstrated that the KZ prediction holds for the disordered Haldane model as long as the spectral gap is not completely destroyed by too strong disorder, and the scaling exponent of the excitation density is robust against the disorder strength. We now turn to the Kane-Mele model, which falls into a different topological class from the Haldane model. 

We again start from the clean limit. We first choose the parameters as $M=0.5$ and $\lambda_{\rm R}=1 $. The spectral gap at half filling with increasing $ \lambda_{\rm S} $ from $0.26$ to $0.41$ is plotted in the inset of Fig.~\ref{3-Sgap_rescaleSgap_Exci_clean_KaneMele}(a). The minimal spectral gap at $ \lambda_{\rm S}\approx 0.34 $ for finite system sizes corresponds to a topological phase transition from the trivial insulating phase to the $Z_2$ topological insulator, which we have confirmed by calculating the spin Chern number. As for the Haldane model, we get estimation $ z\approx 1.004 $ and $ \nu\approx 1 $ from the scaling analysis of the spectral gap, as displayed in Figs.~\ref{3-Sgap_rescaleSgap_Exci_clean_KaneMele}(a) and \ref{3-Sgap_rescaleSgap_Exci_clean_KaneMele}(b). Then the KZ mechanism predicts $n_{\rm ex}\sim\tau^{-1}$. On the other hand, we simulate a quench driven by linearly changing $ \lambda_{\rm S} $ from $ 0.1 $ to $ 0.7 $ (from the trivial to the topological phase). The final excitation density scales with the quench rate as $ n_{\rm ex}\sim\tau^{-1.005}$ for considered system sizes [Fig.~\ref{3-Sgap_rescaleSgap_Exci_clean_KaneMele}(c)], which is in excellent agreement with the KZ prediction. Therefore, the KZ mechanism is valid not only for Chern insulators, but also for $Z_2$ topological insulators in the absence of disorder.

\begin{figure}
	    \centering
         \includegraphics[width=0.98\linewidth]{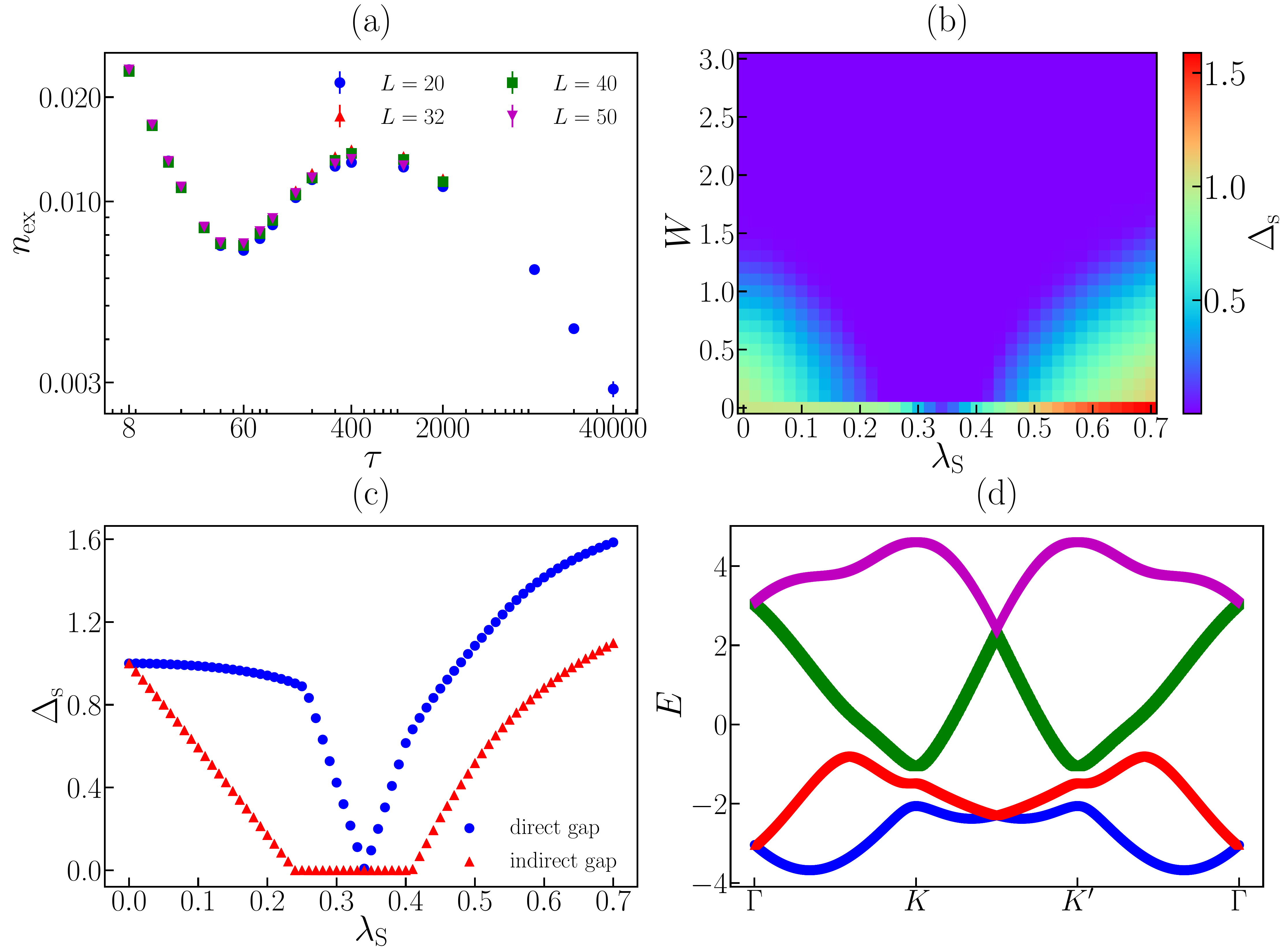}
         \caption{(a) The dependence of the final excitation density $ n_{\rm ex} $ after the linear quench on the quench rate $ 1/\tau $ (on a log-log scale) for the disordered Kane-Mele model with $ W=0.3$. We drive the quench by varying $ \lambda_{\rm S} $ from $ 0.1 $ to $ 0.7 $. $ n_{\rm ex} $ is averaged over 10 disorder configurations at each data point. (b) The spectral gap of the half-filled Kane-Mele model with $ L=50 $ as a function of $ W $ and $ \lambda_{\rm S} $. For $ W=0 $, the gap is closed at a single critical point $ \lambda_{{\rm S}c}\approx0.34 $. The critical point extends to a gapless region when disorder is switched on. This gapless region originates from the vanishing indirect band gap over a finite range of $ \lambda_{\rm S}$ in the clean limit, as shown in panel (c). (d) The band structure of the clean Kane-Mele model along the momentum trajectory $ \Gamma\rightarrow K\rightarrow K^{\prime}\rightarrow\Gamma $ with $ \lambda_{\rm S}=0.3 $, from which the closing of the indirect band gap can be clearly seen. We fix $M=0.5$ and $\lambda_{\rm R}=1 $ in our calculations. }	
	\label{4-Exci_Sgap_EStructure_KaneMele_M_05_W_03}
\end{figure}

\begin{figure*}
	    \centering
         \includegraphics[width=\linewidth]{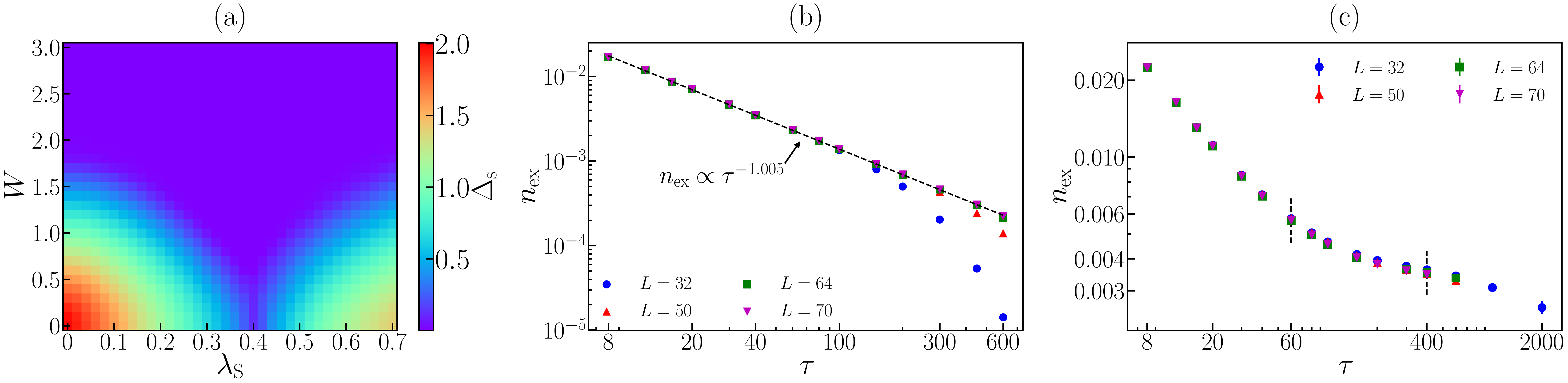}
         \caption{We consider another set of parameter values $M=1 , \lambda_{\rm R}=1 $ for the Kane-Mele model. (a) The spectral gap $ \Delta_{\rm s} $ for $ L=50 $ as a function of $ \lambda_{\rm S} $ and $ W $. The data are averaged over 100 disorder configurations at each parameter point. (b) The dependence of the final excitation density $ n_{\rm ex} $ after the linear quench on the quench rate $ 1/\tau $ (on a log-log scale) for clean systems. For large enough systems, $ n_{\rm ex} $ shows a power-law dependence $ n_{\rm ex}\sim\tau^{-\alpha}$ with $ \alpha\approx 1.005 $. (c) The same quantities as in panel (b), but for the disordered case with $ W=1 $. The dashed lines indicate tentative divisions of three regimes, in which $ n_{\rm ex} $ has different decaying trends. In panels (b) and (c), we drive the quench by varying $ \lambda_{\rm S} $ from 0.1 to 0.7.}	
	\label{5-Sgap_Exci_KaneMele_M_1}
\end{figure*}

We then proceed to the disordered Kane-Mele model. In this case, we find more complicated behavior of the excitation density. In particular, under the same quench from $ \lambda_{\rm S}=0.1 $ to $\lambda_{\rm S}=0.7$, $n_{\rm ex}$ can now depend nonmonotonically with the quench rate. Typical data are shown in Fig.~\ref{4-Exci_Sgap_EStructure_KaneMele_M_05_W_03}(a) for the disorder strength $ W=0.3 $. In this case, $ n_{\rm ex} $ first decays with decreasing quench rate for $ 8\leq\tau\leq60 $, consistent with physical intuition. Surprisingly, it then starts to grow with $\tau$ and reaches a maximum at $ \tau\approx 400 $. Such growth of excitation density at slower quench is the defining feature of the anti-KZ behavior~\cite{PhysRevX.2.041022,antiKZ_PhysRevLett2016,antiKZ_PhysRevB2017,anti_PhysRevA2021,antiKZ_PhysRevB2021,antiKZ_PhysRevLett2020}. At last, $ n_{\rm ex} $ again decays with $ \tau $ for sufficiently slow quench. Even in the first and the third regimes where $ n_{\rm ex} $ decays with increasing $\tau$, we cannot comfirm a power-law decaying of the excitation density consistent with that in Eq.~(\ref{exci_tau}). 

We attribute this nonmonotonic behavior of excitation density to the emergence of a disorder-induced gapless region. Unlike in the disordered Haldane model, the spectral gap of the Kane-Mele model with $M=0.5$ and $\lambda_{\rm R}=1 $ vanishes in a finite range of $\lambda_{\rm S}$ rather than at a single point once disorder is switched on [Fig.~\ref{4-Exci_Sgap_EStructure_KaneMele_M_05_W_03}(b)]. The origin of this gapless region can be understood by tracking the indirect band gap in the clean limit. In Fig.~\ref{4-Exci_Sgap_EStructure_KaneMele_M_05_W_03}(c), we plot the direct and indirect band gaps of the half filled clean Kane-Mele model as functions of $ \lambda_{\rm S} $ for $M=0.5,\lambda_{\rm R}=1 $. The direct band gap is defined as the minimal momentum-resolved energy difference between the conduction and the valence band. By contrast, the indirect band gap measures the energy difference between the conduction-band bottom and the valence-band top, even though they may carry different momenta. One can see that the direct band gap vanishes at a single point of $\lambda_{\rm S}\approx 0.34$. However, the indirect band gap is closed within a finite range of $\lambda_{\rm S}\in[0.24,0.4]$ [Figs.~\ref{4-Exci_Sgap_EStructure_KaneMele_M_05_W_03}(c) and \ref{4-Exci_Sgap_EStructure_KaneMele_M_05_W_03}(d)], which evolves to the gapless region in the presence of disorder. 

In the clean limit, excitations of fermions can only occur within the same momentum sector, so the behavior of $n_{\rm ex}$ is determined by the closing of the spectral gap at the critical point $\lambda_{\rm S}\approx 0.34$, which leads to a similar situation with the Haldane model (Fig.~\ref{2-Sgap_rescaleSgap_Exci_clean_W_1_Haldane}). Nevertheless, fermions are allowed to tunnel between distinct momenta once disorder is switched on, making the indirect band gap matter. In this case, our quench from $\lambda_{\rm S}=0.1$ to $\lambda_{\rm S}=0.7$ drives the system across a gapless region rather than a gapless point, for which the applicability of the KZ theory is not guaranteed. We also observe similar anti-KZ behavior of excitation density at other values of disorder strength using the same model parameters and quench protocol (see Appendix~\ref{sec::moreresults}). 

So far we did not fully understand such anti-KZ behavior of $n_{\rm ex}$ yet, which we believe must depend on the details of nonadiabatic evolution in the gapless region. The nonmonotonic dependence of the excitation density on $\tau$ clearly rules out the possibility that the loss of adiabaticity is dominated only by the critical point on one side of the gapless region~\cite{PhysRevB.77.140404}. In Fig.~\ref{4-Exci_Sgap_EStructure_KaneMele_M_05_W_03}(a), the power-law decaying relation between $n_{\rm ex}$ and $\tau$ could be recovered for extremely slow quench with very large $\tau$. However, due to the expensive numerical simulation of very slow quench, we only reach $\tau=10000, 20000, 40000$ for a small system size $L=20$. We cannot confirm the behavior of $n_{\rm ex}$ in the large $\tau$ limit only based on these data. 

In disordered Kane-Mele model, a gapless region can emerge from other mechanisms. For instance, we have considered another set of model parameters $M=1, \lambda_{\rm R}=1 $. In this case, both the direct and indirect band gaps vanish at a single critical point in the clean limit, which is different from the situation at $M=0.5,\lambda_{\rm R}=1 $. However, a gapless region again appears in the presence of disorder, corresponding to the disorder-induced intermediate metallic phase between the trivial insulator and the $Z_2$ topological insulator~\cite{53-Moore_ChernParity_2007}. The spectral gap as a function of $\lambda_{\rm S}$ and $W$ is displayed in Fig.~\ref{5-Sgap_Exci_KaneMele_M_1}(a) for $L=50$, where the collapse of spectral gap in a finite range of $\lambda_{\rm S}$ becomes clear at $W\approx0.5$. While the excitation density still scales as $\tau^{-1}$ at zero disorder [Fig.~\ref{5-Sgap_Exci_KaneMele_M_1}(b)], quenching across the gapless region when disorder is present destroys the KZ scaling, as shown in Fig.~\ref{5-Sgap_Exci_KaneMele_M_1}(c) for $W=1$. In this case, three regimes with different decaying trends (as reflected by the evolution of the curve's slope) can be identified [Fig.~\ref{5-Sgap_Exci_KaneMele_M_1}(c)]. We also observe such non-KZ decaying of $ n_{\rm ex} $ at other disorder strengths using the same parameters and quench protocol (see Appendix~\ref{sec::moreresults}). 

Although $ n_{\rm ex} $ in Fig.~\ref{5-Sgap_Exci_KaneMele_M_1} violates the KZ prediction, it still monotonically decays with increasing quench rate, without showing the anti-KZ feature. The distinct dependence of $ n_{\rm ex} $ on $\tau$ in Figs.~\ref{4-Exci_Sgap_EStructure_KaneMele_M_05_W_03} and \ref{5-Sgap_Exci_KaneMele_M_1} suggests that whether the anti-KZ (i.e., nonmonotonic) behavior of $ n_{\rm ex} $ appears in disordered Kane-Mele model strongly depends on the mechanism of the formation of the gapless region. The gapless region in Fig.~\ref{4-Exci_Sgap_EStructure_KaneMele_M_05_W_03} originates from the vanishing indirect band gap over a range of $\lambda_{\rm S}$ in the clean limit. Instead, the one in Fig.~\ref{5-Sgap_Exci_KaneMele_M_1} is solely caused by the disorder-induced metallic phase and has nothing to do with the indirect band gap in the clean limit, which vanishes only at a single point of $\lambda_{\rm S}$. While the simple KZ behavior of $ n_{\rm ex} $ is absent in both cases, only the former demonstrates the anti-KZ feature.

\section{Detect the breakdown of adiabatic evolution using particle's occupation}\label{electron occupation}
In previous sections, we have studied the excitation density for the Haldane and Kane-Mele models. However, the excitation density, as defined in Eq.~(\ref{excitation_numerical}), is not convenient for measurement in laboratory. We hence aim to seek an experimentally measurable quantity to detect the breakdown of adiabatic evolution and capture the KZ or other peculiar features in the dynamics. 

Let us first consider clean systems. In the absence of disorder, we can write down the model's Hamiltonian in the momentum space under the sublattice basis. For both the Haldane and Kane-Mele models, the driving terms of the quench in our protocols appear as the diagonal elements of the Hamiltonian matrix expressed in the sublattice basis. This suggests that the momentum-space occupation $\langle n_{{\bm k},\alpha}\rangle$, where $\alpha$ labels the sublattice, may capture the breakdown of the adiabatic evolution. Going back to the real space, as the occupation of particles $\langle n_i\rangle$ on a specific site $i$ in sublattice $\alpha$ is proportional to $ \sum_{\bm k}\langle n_{{\bm k},\alpha}\rangle$, we expect that the site resolved occupation of particles~\cite{PhysRevLett.114.213002}, which is also relevant for the ultracold fermion implementation~\cite{ColdHaldane}, can be used as a local measurable quantity to characterize the quench dynamics. 

To be concrete, for a specific lattice site $i$, we consider the difference $\Delta\rho_i$ between the actual onsite occupation at the end of the quench, obtained from the final time-evolved state, and the static occupation obtained from the final instantaneous ground state. As both the time-evolved state and the instantaneous ground state are Slater determinants of single-particle states, $\Delta\rho_i$ can be expressed as
\begin{eqnarray}\label{DeltaRho}
 \Delta\rho_i=\sum_{m=0}^{N_s/2-1}\left(\langle n_{i}\rangle_{\varphi_m(t=\tau)}-\langle n_{i}\rangle_{\psi_m(\lambda_f)}\right),
\end{eqnarray}
where $|\psi_m(\lambda_f)\rangle$ is the $m$th instantaneous single-particle eigenstate at the final $\lambda=\lambda_f$, and $|\varphi_m(t=\tau)\rangle$ is the final state evolving from the $m$th initial instantaneous single-particle eigenstate $|\psi_m(\lambda_i)\rangle$. It is natural to label these single-particle levels by momentum in the absence of disorder, however, the definition (\ref{DeltaRho}) is still valid in the disordered case. If the system keeps adiabatic evolution, we expect zero $ \Delta\rho$ for any site. Stronger breaking of adiabatic evolution can lead to larger magnitude of $\Delta\rho$. Without loss of generality, in the following we assume the site $i$ belongs to the $A$ sublattice of the honeycomb. For the sake of measurement convenience, we sum over both spins on the site for the Kane-Mele model (we have examined that each individual spin gives similar results). Because $\Delta\rho$ could be negative, we display its absolute value. 

\begin{figure}
	    \centering
         \includegraphics[width=0.98\linewidth]{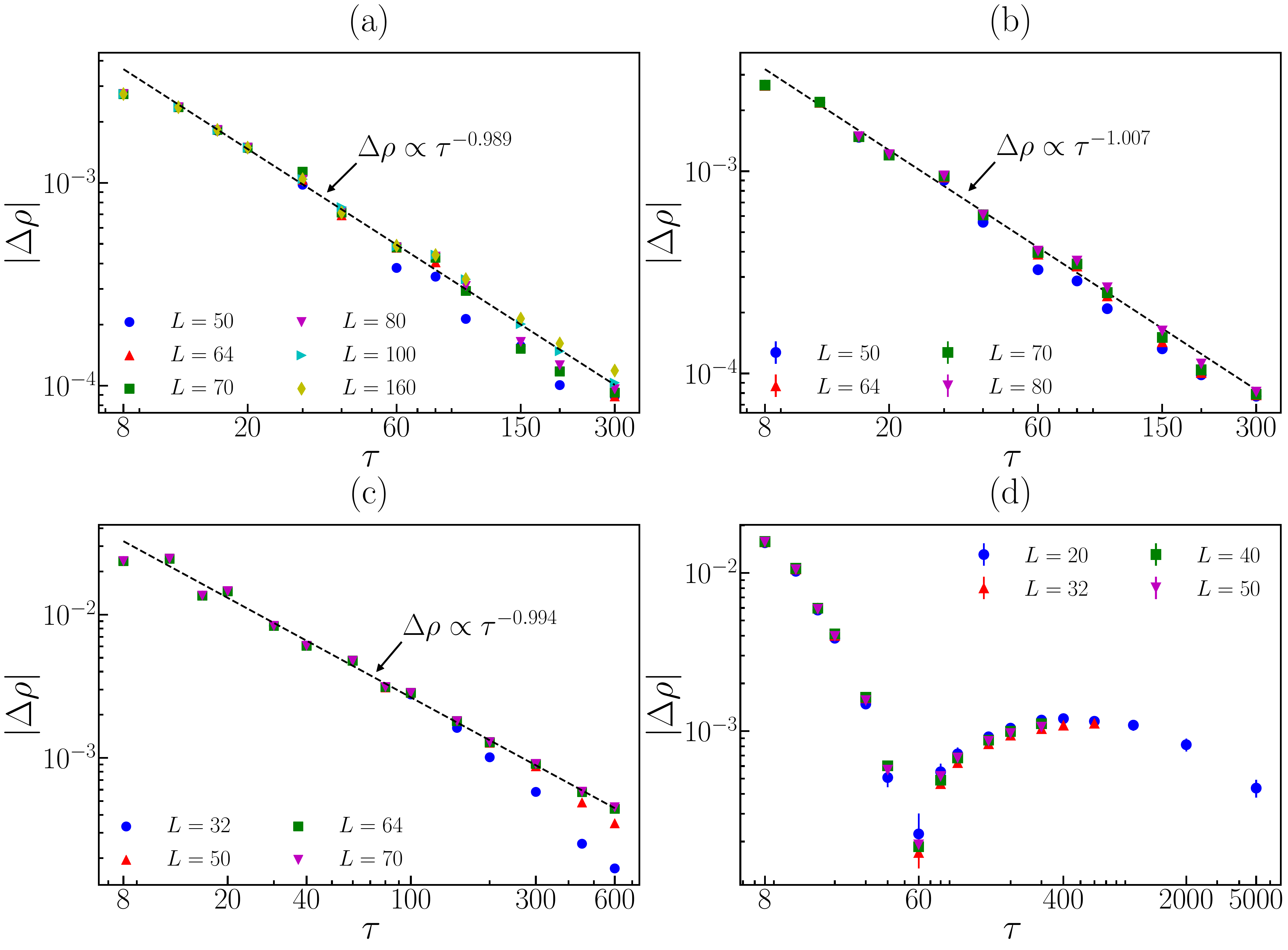}
         \caption{The particle's occupation difference $\Delta\rho$ as a function of quench rate $1/\tau$ (on a log-log scale) for (a) the clean Haldane model, (b) the disordered Haldane model with $ W=1 $, (c) the clean Kane-Mele model, and (d) the disordered Kane-Mele model with $W=1$. Model parameters and quench protocols are the same as those in Figs.~\ref{2-Sgap_rescaleSgap_Exci_clean_W_1_Haldane} and \ref{5-Sgap_Exci_KaneMele_M_1}. In panels (a), (b), and (c), $\Delta\rho$ shows a power-law decay $\Delta\rho\sim\tau^{-\alpha}$ for sufficiently large systems, with the exponent $\alpha\approx 0.989 $, $ 1.007$, and $ 0.994 $, respectively. In panel (d), the behavior of $|\Delta\rho|$ unveils the collapse of the KZ scaling.}
          \label{6-Occup_Haldane_KaneMele}
\end{figure}
 
Remarkably, $ \Delta\rho$ indeed exhibits very similar behavior with the excitation density for clean systems under PBCs, as displayed in Figs.~\ref{6-Occup_Haldane_KaneMele}(a) and \ref{6-Occup_Haldane_KaneMele}(c) for the Haldane and Kane-Mele models, respectively. In this case, the occupation difference $ \Delta\rho_i$ does not depend on in which unit cell the site $i$ is located. It keeps positive and shows a power-law decay with increasing $\tau$. The numerically extracted exponents are very close to $1$, in good agreement with the KZ prediction for the excitation density. 

While our choice of $ \Delta\rho$ is motivated by the analysis in the clean limit, we find that it can detect the breakdown of adiabatic evolution even when disorder is present. To reduce the number of required samples, we average $ \Delta\rho_i$ over the $A$ sites of all unit cells for each disorder configuration. For the disordered Haldane model, $ \Delta\rho$ again behaves similarly to the excitation density, with the exponent of the power-law decaying close to the KZ prediction [Fig.~\ref{6-Occup_Haldane_KaneMele}(b)]. By contrast, we observe the anti-KZ behavior of $ \Delta\rho$ for the disordered Kane-Mele model with $M=1 , \lambda_{\rm R}=1 $ [Fig.~\ref{6-Occup_Haldane_KaneMele}(d)]. At first $ \Delta\rho$ is positive and decreases with increasing $\tau$, consistent with the physical intuition that the breaking of adiabatic evolution becomes weaker for slower quench. However,  in the range of $ 60\leq\tau\leq400 $, $ \Delta\rho $ becomes negative and its absolute value increases with $\tau$, indicating the unexpected stronger breaking of adiabatic evolution at smaller quench rate. Finally, at larger $\tau$, $ \Delta\rho $ keeps negative but its absolute value decays, consistent with the physical intuition again. While the excitation density itself in this case does not show the anti-KZ behavior, the turning points $\tau=60$ and $\tau=400$ in $ |\Delta\rho| $ capture the change of $ n_{\rm ex} $'s decaying trend in Fig.~\ref{5-Sgap_Exci_KaneMele_M_1}(c) well. We find similar results for other disorder strengths (see Appendix~\ref{sec::moreresults}). We have also considered the another set of parameters $M=0.5,\lambda_{\rm R}=1 $ as in Fig.~\ref{4-Exci_Sgap_EStructure_KaneMele_M_05_W_03} for the disordered Kane-Mele model. In this case both $ n_{\rm ex} $ and $ |\Delta\rho| $ demonstrate the anti-KZ behavior with $\tau$, and their turning points match each other (see Appendix~\ref{sec::moreresults}). We hence conclude that $ |\Delta\rho| $ is sufficiently sensitive to detect not only the anti-KZ feature of $n_{\rm ex}$, but also other behavior transitions of $n_{\rm ex}$.

\begin{figure}
	    \centering
         \includegraphics[width=0.98\linewidth]{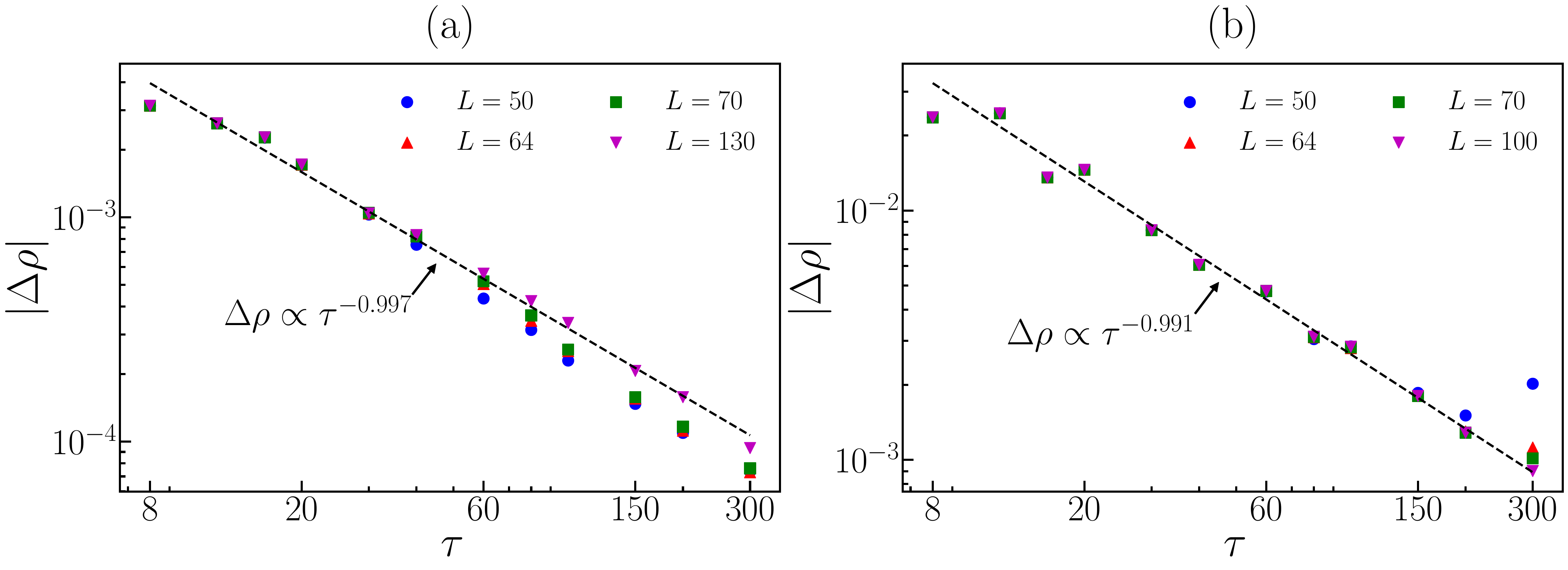}
         \caption{The particle's occupation difference $\Delta\rho$ as a function of quench rate $1/\tau$ (on a log-log scale) for (a) the clean Haldane model and (b) the clean Kane-Mele model under open boundary conditions. We display $\Delta\rho$ on the $A$ site of the central unit cell of each finite system, which shows a power-law relation $\Delta\rho\sim\tau^{-\alpha}$ with $\alpha\approx 0.997 $ and $ 0.991 $, respectively. Model parameters and quench protocols are the same as those in Figs.~\ref{2-Sgap_rescaleSgap_Exci_clean_W_1_Haldane} and \ref{5-Sgap_Exci_KaneMele_M_1}.}
          \label{7-Occup_Haldane_KaneMele_OBC}
 \end{figure}
 
Finally, considering that it is difficult to realize 2D periodic systems in experiments, we further study the behavior of $ \Delta\rho_i$ under OBCs. Unlike the case with PBCs, the presence of gapless edge states, which is a defining feature of topological band insulators, can significantly complicate the dynamics. For instance, the adiabatic evolution is broken by the gapless edge states, even when the quench is constrained to a topological phase where the bulk gap remains open. Furthermore, the excitation density can exhibit anti-KZ behavior when a system with open boundaries is driven across a single critical point of a topological phase transition~\cite{52-KunYang_PRB_2018}. Surprisingly, we find that $\Delta\rho_i$ in the bulk of the system under OBCs have very similar behavior to that under PBCs, despite the complexity due to edge states. The data 
of the clean Haldane and Kane-Mele model are displayed in Figs.~\ref{7-Occup_Haldane_KaneMele_OBC}(a) and \ref{7-Occup_Haldane_KaneMele_OBC}(b), respectively, in which $\Delta\rho_i$ is computed on the $A$ site of the unit cell at the center of the system. $\Delta\rho$ demonstrates power-law decaying with increasing $\tau$ for both models, and the  corresponding powers are in excellent agreement with those under the PBCs. 

The results in this section indicate that the occupation difference defined in Eq.~(\ref{DeltaRho}) is a promising measurable quantity for the experimental characterization of linear quench dynamics in the Haldane and Kane-Mele models. We also expect similar results for other models, if the driving parameters of the quench appear in the diagonal elements of the Hamiltonian in the sublattice basis. However, we should emphasize that Eq.~(\ref{DeltaRho}) may fail to capture the dynamics if this condition is not satisfied. In this case, it is necessary to modify Eq.~(\ref{DeltaRho}), which may introduce crossing terms between different lattice sites (see Appendix~\ref{sec::checkerboard} for an example). These crossing terms break the onsite locality and complicate experimental measurement.

\section{Conclusions and outlook}\label{c_and_o}
In this work, we study the finite-rate quench dynamics in the Haldane and Kane-Mele model. The quench is driven by varying a specific model parameter from a trivial insulating phase to a topological band insulator phase. We characterize the breakdown of adiabatic evolution by properly defining the excitation density in the absence of symmetry breaking. For both the clean and disordered Haldane models, where the pertinent topological phase transition is characterized by a single critical point, we find the power-law decaying of the excitation density with decreasing quench rate consistent with the Kibble-Zurek prediction, as long as the spectral gap is not closed by too strong disorder. For the Kane-Mele model involving the $Z_2$ topological phase, while the KZ behavior of excitation density is also observed in the clean limit, disorder can destroy the KZ scaling by extending the single critical point to a gapless region and even induce the anti-KZ feature. We propose using the fermion's onsite occupation to experimentally detect the breakdown of the adiabatic evolution and capture the KZ and anti-KZ behavior of the excitation density. The locality of the onsite occupation facilitates its application even for realistic systems with physical edges.

There are a couple of open questions which deserve future investigations. For the disordered Kane-Mele model, it would be nice if we can incorporate disorder in the Landau-Zener analysis to study the scattering of fermions between different momenta across the indirect band gap. This could be helpful for understanding the anti-KZ dependence of excitation density on the quench rate at weak disorder. Second, there are other quantities that can characterize the breakdown of adiabatic evolution, such as the energy and entropy production~\cite{RevModPhys.83.863}. It would be interesting to study the behaviors of these quantities in our models. Moreover, we focus on noninteracting two-dimensional Chern and $Z_2$ topological models in this work. It is attractive to study finite-rate quench across quantum phase transitions and disorder effects for topological systems in other spatial dimensions and/or in other topological classes, as well as for interacting systems.  At last, we consider uniform global quench in this work. How to extend these results to local or spatially nonuniform quench remains an open question.

\acknowledgements
We appreciate Jernej Mravlje, Ferdinand Evers, Jan Carl Budich, and Anders Sandvik for helpful discussions. We gratefully thank Chao Cao for GPU resources, which we uesd to run part of numerical simulations. This project was supported by the National Key Research and Development Program of China (Grant No.~2021YFA1401902) and the National Natural Science Foundation of China (Grant No.~11974014).

\section*{Data availability}\label{data}
The data that support the findings of this article are not publicly available. The data are available from the authors upon reasonable request.

\appendix

\section{Behavior of excitation density after deviating the KZ prediction}\label{sec::largetau}
In the main text, we discussed a finite-size effect for the Haldane model, in which the excitation density deviates from the $\tau^{-1}$ scaling for sufficiently large $\tau$. This deviation, due to the nonzero energy gap of a finite system, occurs at smaller $\tau$ for smaller systems. Applying the adiabatic perturbation theory in the large $\tau$ limit for a finite system leads to the prediction of the $ \tau^{-2} $ scaling of excitation density~\cite{PhysRevB.81.012303}. 

\begin{figure}
	    \centering
         \includegraphics[width=0.98\linewidth]{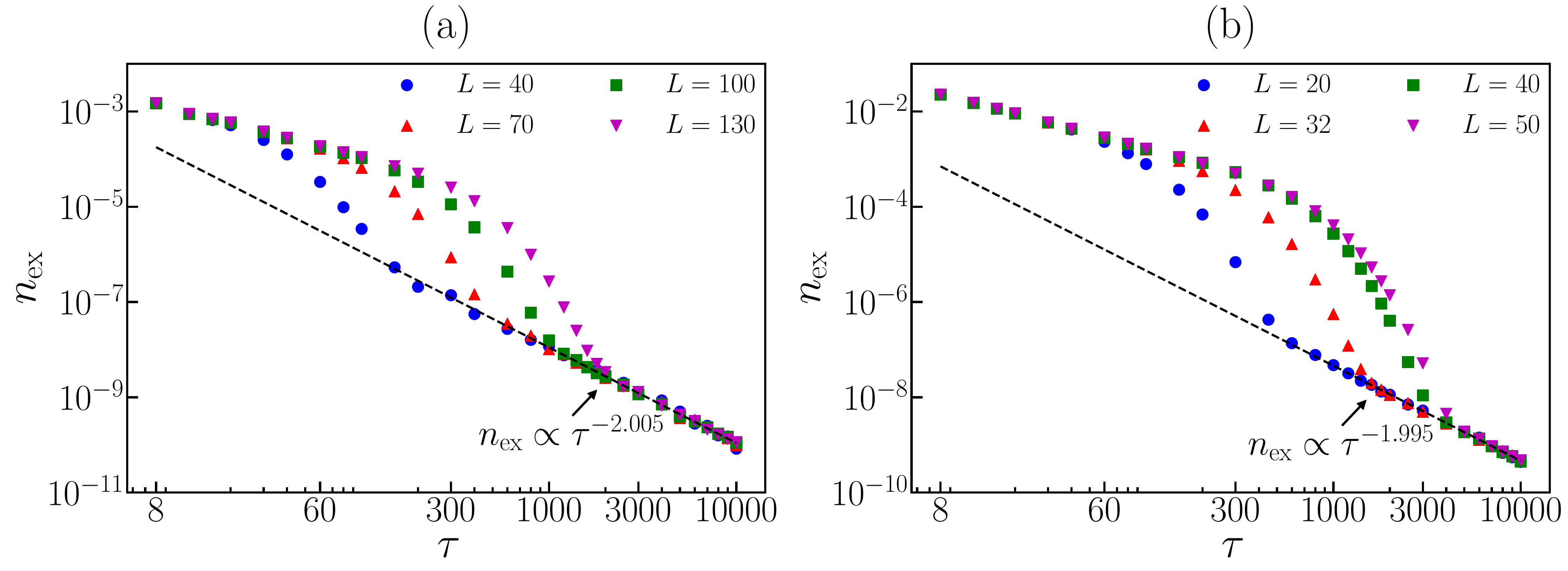}
         \caption{The dependence of the final excitation density $ n_{\rm ex} $ on the quench rate $ 1/\tau $ (on a log-log scale) for (a) clean Haldane model and (b) clean Kane-Mele model.  We consider $\tau$ up to $10^4$ to reach very slow quench. The $ \tau^{-2} $ scaling of the excitation density can be observed for both models at sufficiently large $\tau$.}
          \label{AppxA-Exci_very_large_tau}
\end{figure}

Now we numerically examine this for the clean Haldane and Kane-Mele models under PBCs. We choose $ M=1, \phi=\pi/2 , t_{3}=1/3 $ for the Haldane model and $ M=0.5 , \lambda_{\rm R}=1 $ for the Kane-Mele model. The quench protocols are the same as those in Sec.~\ref{disorder effect}. As shown in Fig.~\ref{AppxA-Exci_very_large_tau}, for each finite system there is a clear transition of the excitation density at the end of the quench from the $\tau^{-1}$ scaling to the $\tau^{-2}$ scaling with the increasing of $\tau$. This transition happens at larger $\tau$ for bigger systems. We expect that only the $\tau^{-1}$ scaling is present in the thermodynamic limit.

\section{More results of disordered Kane-Mele model}\label{sec::moreresults}
In the main text, we have studied the finite-rate quench dynamics for disordered Kane-Mele model using two sets of model parameters, $M=0.5 , \lambda_{\rm R}=1 $ and $M=1 , \lambda_{\rm R}=1 $, with disorder strengths $W=0.3$ and $W=1$, respectively [see Figs.~\ref{4-Exci_Sgap_EStructure_KaneMele_M_05_W_03}(a), \ref{5-Sgap_Exci_KaneMele_M_1}(c), and \ref{6-Occup_Haldane_KaneMele}(d)]. Here we present more numerical data obtained with different disorder strengths.

\begin{figure}
	    \centering
         \includegraphics[width=0.98\linewidth]{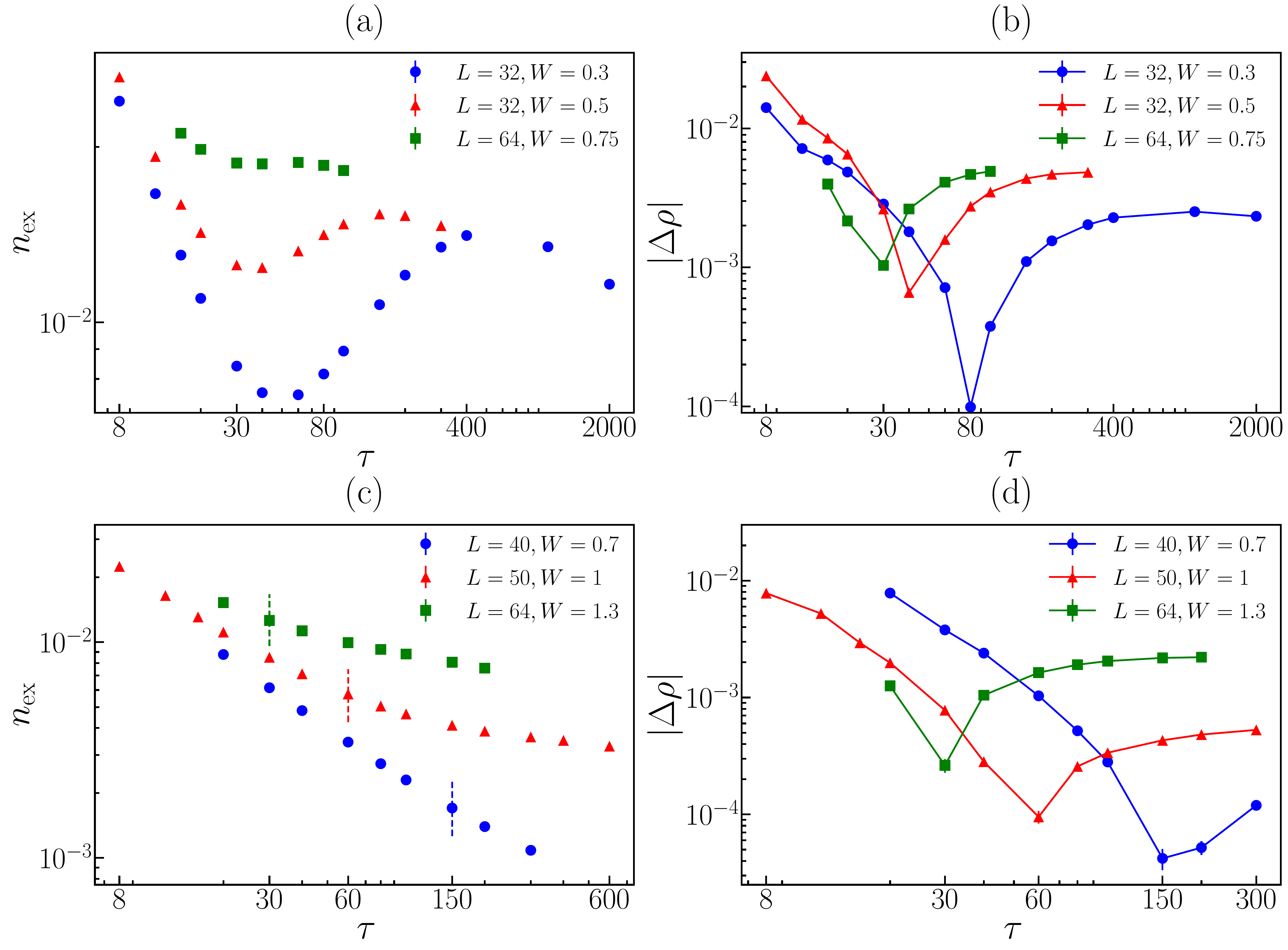}
         \caption{(a) The dependence of the final excitation density $ n_{\rm ex} $ and (b) the particle's occupation difference $\Delta\rho$ after the linear quench on the quench rate $1/\tau$ (on a log-log scale) for disordered Kane-Mele model with $ M=0.5 $. Both of the $ n_{\rm ex} $ and $\Delta\rho$ show anti-KZ feature with quench rate for different disorder strengths. (c) and (d) The same quantities as in panels (a) and (b), but for $ M=1 $. The dashed lines in panel (c) indicate the turning points of $|\Delta\rho|$ in panel (d), which can be used as tentative divisions of regimes in which $ n_{\rm ex} $ has different decaying trends (as reflected by the evolution of the curve's slope). We drive the quench by varying $ \lambda_{\rm S} $ from 0.1 to 0.7.}
          \label{AppxC-Exci-Occup-KaneMele_M_0.5-1}
\end{figure}

For $M=0.5 , \lambda_{\rm R}=1 $, a gapless region exists in disordered systems. It originates from the vanishing indirect band gap over a range of quench parameter $\lambda_{\rm S}$ in the clean limit, which turns to the gapless region once particles are allowed by disorder to hop between different momenta. When varying $\lambda_{\rm S}$ across this region, we persistently observe the anti-KZ behavior of excitation density $ n_{\rm ex} $ at various disorder strengths, as shown in Fig.~\ref{AppxC-Exci-Occup-KaneMele_M_0.5-1}(a) for $W=0.3$, $0.5$, and $0.75$. We also notice that larger system sizes are needed to get an obvious anti-KZ feature for stronger $W$: $L=32$ is sufficient for $W=0.3$ and $0.5$, but we have to reach $L=64$ for $W=0.75$. The particle's occupation difference $\Delta\rho$ exhibits a similar anti-KZ behavior with $ \tau $ [Fig.~\ref{AppxC-Exci-Occup-KaneMele_M_0.5-1}(b)]. One can see that the turning points of $|\Delta\rho|$ and $n_{\rm ex}$ match each other very well.

For $M=1 , \lambda_{\rm R}=1 $, a gapless region also exists in disordered systems, but its origin is different from that at $M=0.5 , \lambda_{\rm R}=1 $. In this case, the indirect band gap vanishes at a single critical point $ \lambda_{{\rm S}c} $ together with the direct band gap in the absence of disorder, thus being irrelevant with the emergence of a gapless region. The gapless region now actually corresponds to the metallic phase induced by disorder in two-dimensional time-reversal-invariant systems with spin-orbit coupling~\cite{53-Moore_ChernParity_2007}. In the main text, we find there is no anti-KZ feature of the excitation density although it does not obey the KZ scaling. Similar results are observed for various disorder strengths, as shown in Fig.~\ref{AppxC-Exci-Occup-KaneMele_M_0.5-1}(c). However, the particle's occupation difference $\Delta\rho$ does exhibit anti-KZ behavior with $ \tau $ [Fig.~\ref{AppxC-Exci-Occup-KaneMele_M_0.5-1}(d)]. Comparing Figs.~\ref{AppxC-Exci-Occup-KaneMele_M_0.5-1}(c) and \ref{AppxC-Exci-Occup-KaneMele_M_0.5-1}(d), one can see that the turning points of $|\Delta\rho|$ reflects the change of $n_{\rm ex}$'s decaying trend at all disorder strengths, consistent with the results in the main text.

Our results bring out two important messages: (i) whether the quench across a gapless region leads to the anti-KZ behavior of excitation density strongly depends on the mechanism of the formation of the gapless region; and (ii) the particle's occupation difference is a sensitive quantity to the behavior transition of excitation density with respect to the quench rate. Such behavior transition may correspond to either the change of decaying trend or the anti-KZ feature. A further study is needed to understand why $|\Delta\rho|$ can still show anti-KZ behavior even when excitation density itself consistently decays with increasing $\tau$.

\section{Particle's onsite occupation for the checkerboard model}
\label{sec::checkerboard}
In the main text, we have shown that particle's onsite occupation difference Eq.~(\ref{DeltaRho}) can be used to detect the breakdown of adiabatic evolution in Haldane and Kane-Mele models. We expect similar results for other models, if the driving parameters of the quench appear in the diagonal elements of the Hamiltonian in the sublattice basis. Now we consider a model for which this condition is not satisfied. 

We consider the checkerboard model with two square sublattices of lattice constant $a$. Including real isotropic NN hoppings, purely imaginary NN hoppings that depend on the hopping direction, and two kinds of real NNN hoppings in orthogonal directions, the single-particle Hamiltonian is~\cite{52-KunYang_PRB_2018} 
\begin{eqnarray}\label{checkerboard_real}
H = &-&\sum_{\bm{r},\bm{\delta}}c_{{\bm r},A}^{\dagger}c_{\bm{r}+\bm{\delta},B}+iV\sum_{{\bm r},\bm{\delta}}D_{\bm{\delta}}c_{\bm{r},A}^{\dagger} c_{\bm{r}+\bm{\delta},B} + {\rm H.c.}\nonumber \\
&-& \xi\sum_{\bm{r}}(c_{\bm{r},A}^{\dagger}c_{\bm{r}\pm\bm{a}_{1},A}+c_{\bm{r},B}^{\dagger}c_{\bm{r}\pm\bm{a}_{2},B}) \nonumber \\
&-& \xi^{\prime}\sum_{\bm{r}}(c_{\bm{r},A}^{\dagger}c_{\bm{r}\pm\bm{a}_{2},A}+c_{\bm{r},B}^{\dagger}c_{\bm{r}\pm\bm{a}_{1},B}),
\end{eqnarray}
where $c_{{\bm r},A/B}^\dagger$ creates a fermion on the site of sublattice $A/B$ at position ${\bm r}$, $ \bm{a}_{1}=(0,a) $, $ \bm{a}_{2}=(a,0) $, and $ D_{\bm{\delta}}=1 $ if $ \bm{\delta}=\pm(a/2,a/2) $ and $ D_{\bm{\delta}}=-1 $ if $ \bm{\delta}=\pm(a/2,-a/2) $. We neglect disorder here. We choose $\xi=-\xi'=0.5$ and set $a=1$ in what follows. 
\begin{figure}
	    \centering
         \includegraphics[width=0.98\linewidth]{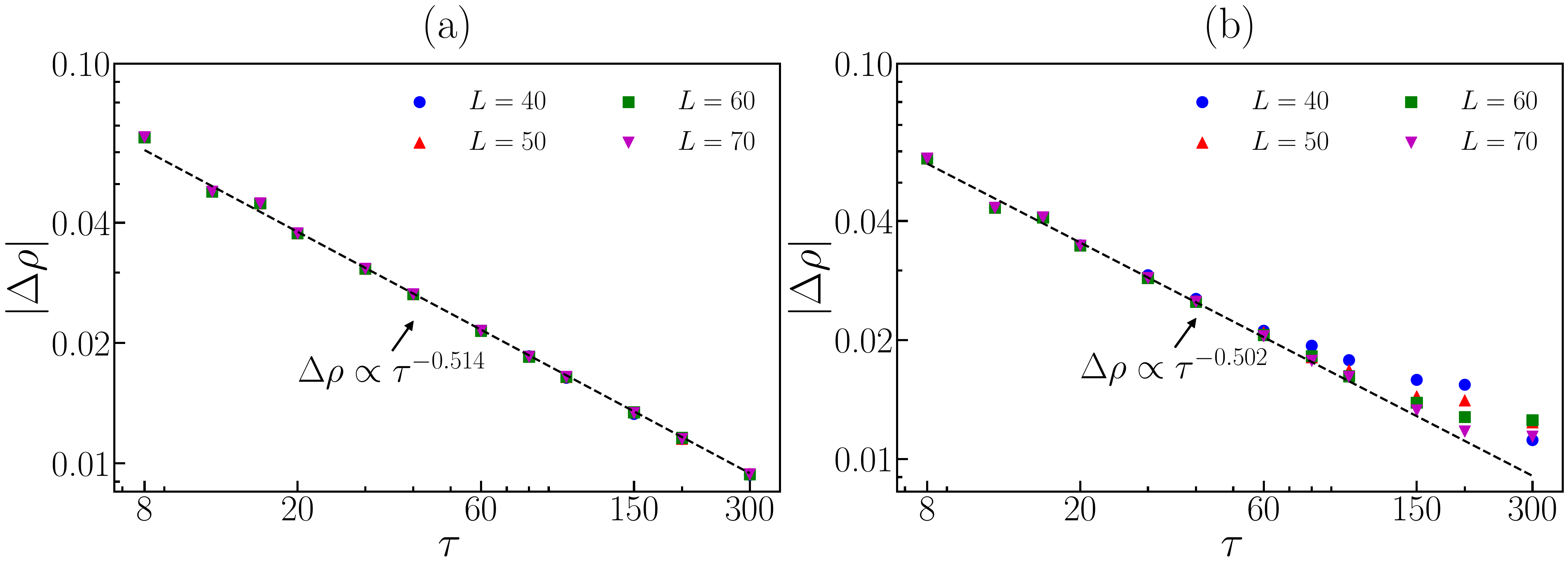}
         \caption{The modified particle's occupation difference $ {\Delta\rho} $ as a function of quench rate $ 1/\tau $ (on a log-log scale) for the clean checkerboard model under (a) periodical boundary conditions and (b) open boundary conditions. }
          \label{AppxB-Occup_checkerboard_PBC_OBC_clean}
\end{figure}

We drive the quench by varying the imaginary NN hopping parameter $ V $ from $ 1 $ to $ -1 $. The system goes across a topological phase transition at $V=0$, with the Chern number of the lower band changing from $ -1 $ to $ 1 $, which appears in the off-diagonal element of the Hamiltonian under the sublattice basis. As we know $z=2$ and $\nu=1$ for this transition, the KZ mechanism predicts the $ \tau^{-0.5}$ scaling for the excitation density at the end of the quench. The band gap is closed at $ {\bm{k}}=(\pi, \pi) $ when the transition happens. The low-energy Hamiltonian describing the transition is 
\begin{equation}\label{checkerboard_K}
H(\bm{q}) = q_x q_y\sigma_x+4V\sigma_y-\frac{1}{2}(q_x^2-q_y^2)\sigma_z
\end{equation}
in the sublattice basis, where ${\bm q}$ is the deviation from $(\pi,\pi)$ and $\sigma$'s are Pauli matrices. Since the driving parameter $V$ appears in the off-diagonal elements, this Hamiltonian does not satisfy the initial condition of the standard Landau-Zener theory, which requires that one level is fully occupied and the other is empty. To use the LZ theory, one needs to apply a unitary transformation 
\begin{equation}
U = \frac{1}{\sqrt{2}}
\left(
\begin{array}{cc}
   1&i \\
   1 & -i
\end{array}
\right),
\end{equation}
such that
\begin{equation}\label{checkerboard_K_unitary}
UH({\bm{q}})U^\dagger = 
\left(
\begin{array}{cc}
  -4V & -\frac{1}{2}(q_x-iq_y)^{2} \\
   -\frac{1}{2}(q_x+iq_y)^{2} & 4V
\end{array}
\right).
\end{equation}
In this transformed Hamiltonian, the quadratic dependence of the off-diagonal elements on momentum gives the quadratic band structure near the band-touching point. Application of the LZ theory on this transformed Hamiltonian gives $n_{\rm ex}\propto \tau^{-0.5}$, which matches the KZ prediction and was numerically confirmed in Ref.~\cite{52-KunYang_PRB_2018}.

The necessity of the unitary transformation implies that we need to modify the definition of particle's occupation accordingly if we want to observe the $\tau^{-0.5}$ scaling in $\Delta\rho$. For instance, for a specific $A$ site, $\langle n_A\rangle$ in Eq.~(\ref{DeltaRho}) should be replaced with $\langle c_A^\dagger c_A+c_B^\dagger c_B+ic_A^\dagger c_B-ic_B^\dagger c_A\rangle/2$.
We have computed such modified occupation difference for the clean checkerboard model under either PBCs or OBCs. As shown in Fig.~\ref{AppxB-Occup_checkerboard_PBC_OBC_clean}, we indeed observe clear $\tau^{-0.5}$ scaling in both cases. By contrast, we find that the original occupation difference Eq.~(\ref{DeltaRho}) keeps almost zero for any $\tau$. 
 
\bibliography{SlowQuench}

\end{document}